\documentclass[
    ,final            
  ]
  {aipproc}

\layoutstyle{6x9}

\begin{document}

\title{Thermal model at RHIC, part II: elliptic flow and HBT radii
\thanks{Talk presented at II International Workshop on Hadron Physics,
{\em Effective Theories of Low-Energy QCD}, 25-29 September 2002,
University of Coimbra, Portugal}}

\author{Wojciech Broniowski, Anna Baran, and Wojciech Florkowski}
{address={The H. Niewodnicza\'nski Institute of Nuclear Physics,
PL-31342 Cracow, Poland}}

\begin{abstract}
We continue the analysis of the preceding talk with a 
discussion of the elliptic flow and the 
Hanbury-Brown--Twiss pion correlation radii. It is shown that the thermal model 
can be extended to describe these phenomena. The description of the 
elliptic flow involves an appropriate deformation of the freeze-out 
hyper-surface and flow velocity. The obtained results agree reasonably with 
the data for soft ($< 2$~GeV) transverse momenta. 
For the pionic HBT correlation radii the experimental 
feature that $R_{\rm out}/R_{\rm side} \simeq 1$ is naturally obtained. 
The reproduction of individual $R_{\rm side}$ and $R_{\rm out}$ 
can be achieved with the inclusion of the excluded volume corrections, which 
effectively increase the radii by $\sim 30\%$.
\end{abstract}

\maketitle

\section{Introduction}

In the preceding talk \cite{prec}, from now on referred to as (I), 
it has been shown that the thermal approach 
is successful in the description of the particle ratios and $p_\perp$-spectra 
at RHIC. Here we continue our investigation, studying azimuthal asymmetry of the 
spectra and the pionic Hanbury-Brown--Twiss correlation radii.

\section{Elliptic flow}

When the nuclei collide at non-zero impact parameter, $b \neq 0$, the momentum distribution of the
produced particles carries azimuthal asymmetry. 
In general, at mid-rapidity ($y=0$) we may write the following Fourier decomposition in 
the azimuthal angle $\phi$, measured from the reaction plane: 
\begin{eqnarray}
\left . \frac{dN}{d^2p_\perp dy} \right |_{y=0}=
\left . \frac{dN}{2\pi p_\perp dp_\perp dy} \right |_{y=0} 
\left ( 1+ 2v_2 \cos 2\phi + 2v_4 \cos 4\phi + \dots \right ).
\end{eqnarray}
The sines are absent due to the symmetry condition $\phi \to -\phi$, which is simply the reflexion with
respect to the reaction plane, while the coefficients of 
cosines with odd multiples of $\phi$ vanish for the case 
of symmetric nuclei and at $y=0$, when the symmetry $\phi \to \pi - \phi$ holds. 
The elliptic-flow coefficient, $v_2$, can therefore be computed as    
\begin{eqnarray}
v_2=\frac{\int_0^{2\pi}\left . \frac{dN}{d^2p_\perp dy}\right | _{y=0} 
\cos 2\phi \, d\phi }{\int_0^{2\pi} \left . \frac{dN}{d^2p_\perp dy} \right | _{y=0} \, d\phi }.
\label{v2def}
\end{eqnarray}

The value of $v_2$ is an important signature of the physics occurring in heavy-ion collisions. 
Most importantly, its non-vanishing value indicates that the production mechanism is not a simple 
composition of nucleon-nucleon collisions, since in that case the asymmetry of production in each
such collision would average out practically to zero. Thus, interactions with other particles 
(rescattering, asymmetric collective flow, \dots ) are necessary 
to generate non-vanishing $v_2$. In hydrodynamical
approaches the elliptic flow has been analyzed in many papers, see {\em e.g.}
\cite{Kolb:1999es,Voloshin:1999gs,Kolb:2000fh,Hirano:2000eu,Magas:2000qs}). 
The coefficient $v_2$ depends on the impact parameter, $b$, on the transverse momentum $p_\perp$, 
as well as, obviously, on the type of the considered particle. All these dependences are measured at RHIC. 
The impact parameter, $b$, is traded for the experimentally more useful 
centrality parameter, $c$, which to a very good accuracy 
is given by \cite{central} 
\begin{eqnarray}
c \simeq \frac{b^2}{(2R)^2}.
\end{eqnarray}  

There are two basic empirical facts from RHIC which we will use in our approach. Firstly,
$v_2$ {\em is positive} 
\cite{Adler:2001zd,Ackermann:2000tr,Snellings:2001nf,Lacey:2001va,%
Adler:2001nb,Adler:2002pu,Adcox:2000sp,Adcox:2001mf,Adcox:2002uc,Adcox:2002ms}, 
which means that the collective flow is faster in the reaction plane than out-of-plane. 
Secondly, the measurement \cite{Snellings:2001fs} 
of the azimuthal dependence of the $R_{\rm side}$ pion HBT radius
shows, that the shape of the system at freeze-out {\em is elongated out of the reaction plane}.
We will now use these two facts in our choice of the 
parameterization of the hypersurface at freeze-out.  
A natural extension of Eq. (I.6) is to introduce the azimuthal shape asymmetry,
\begin{eqnarray}
r_x&=&\rho_{\rm max} \sqrt{1-\epsilon} \cos \phi , \nonumber \\
r_y&=&\rho_{\rm max}  \sqrt{1+\epsilon} \sin \phi , \label{rmod}
\end{eqnarray}
with $r_z$ and $t$ kept as in the symmetric case of Eq. (I.6) of paper (I). 
Our convention is that $r_x$ lies in the reaction plane, and $r_y$ is perpendicular to the reaction plane.
For positive $\epsilon$ this produces elongation
out of the reaction plane, as seen in the experiment. 
The four-velocity of Eq. (I.3) is modified as follows: 
\begin{eqnarray}
u_x&=& \frac1N r_x \sqrt{1+\delta} \cos \phi, \nonumber \\
u_y&=& \frac1N r_y \sqrt{1-\delta} \sin \phi, \nonumber \\
u_z&=& \frac1N r_z, \nonumber \\
u_t&=& \frac1N t . \label{umod}
\end{eqnarray}
The normalization $N$ is such that $u^\mu u_\mu=1$. Positive $\delta$ means faster flow in the reaction plane, which 
corresponds to positive $v_2$. Certainly, the choice (\ref{rmod},\ref{umod}) is by no means unique, 
but it grasps the essential empirical features.

\begin{figure}[tb]
\includegraphics[width=11.5cm]{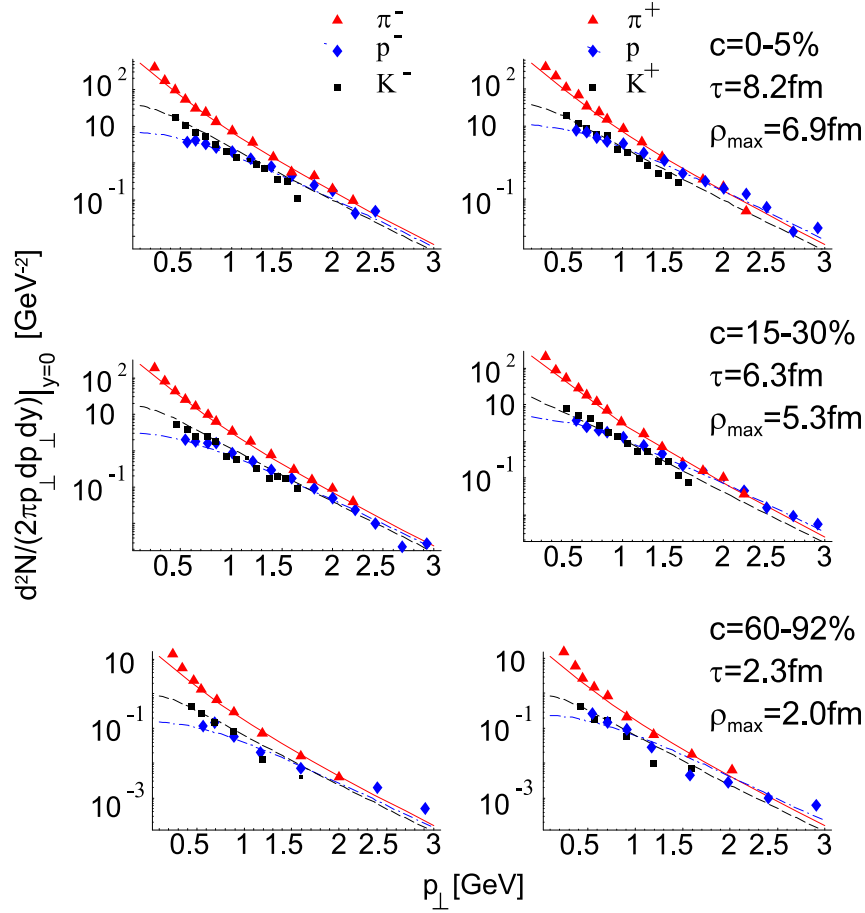} 
\caption{The model fit of the pion, kaon, and proton spectra to the PHENIX data 
for $\sqrt{s_{NN}}=130~{\rm GeV}$ \cite{Adcox:2001mf} 
at three centrality bins, arranged top to bottom. 
Negative (positive) hadron are shown in the left (right) side. The optimum values of the size parameters 
$\tau$ and $\rho_{\rm max}$ are given for each centrality bin.}
\label{fig:allp}
\end{figure}
\begin{table}[b]
\begin{tabular}{lrrrrr}
\hline
& \multicolumn{4}{c}{PHENIX @130GeV} & PHENIX+STAR  \\
& \multicolumn{4}{c}{} & @130GeV  \\
 $c$ [\%]   & min. bias & 0-5 & 15-30 & 60-92 & 0-5/0-6 \\
\hline
$\tau$ [fm]           & 5.6 & 8.2 & 6.3 & 2.3 & 7.7 \\
$\rho_{\rm max}$ [fm] & 4.5 & 6.9 & 5.3 & 2.0 & 6.7 \\
\hline
$\rho_{\rm max}/\tau$ &   0.81   & 0.84 & 0.84  & 0.87 & 0.87  \\
$\beta_\perp^{\rm max}$ &  0.62  &0.64 & 0.64 & 0.66 & 0.66 \\
$ \langle \beta_\perp \rangle$  & 0.46 & 0.47 & 0.47 & 0.48 & 0.48 \\
\hline
\end{tabular}
\label{tab:par}
\caption{Comparison of the size parameters obtained by fitting particle spectra at various 
values of th centrality parameter $c$. Their ratio, as well as the maximum and average value 
of the flow parameter, $\beta_\perp$, are also given.}
\end{table}

\begin{figure}[tb]
\includegraphics[width=11cm]{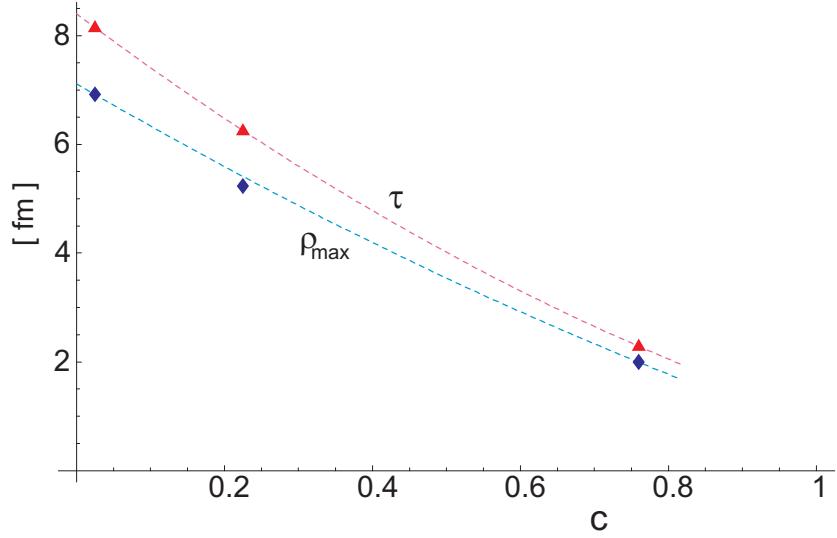} 
\caption{The dependence of the size parameters $\tau$ and $\rho_{\rm max}$ on the centrality 
parameter $c$, as fitted to the PHENIX data on $\phi$-integrated $p_\perp$-spectra 
at $\sqrt{s_{NN}}=130~{\rm GeV}$ \cite{Adcox:2001mf}.}
\label{fig:rt}
\end{figure}

The modified expansion model has four geometric parameters: $\tau$, $\rho_{\rm max}$, $\epsilon$ and $\delta$. 
These parameters depend on the centrality parameter, $c$.
Fortunately, the effect of $\epsilon$ and $\delta$ on the $\phi$-averaged spectra, 
$\left . {dN}/({2\pi p_\perp dp_\perp dy}) \right |_{y=0}$, is negligible and enters at the level
of a few percent. Thus we may first fit $\tau$ and $\rho_{\rm max}$ to the  
$\phi$-averaged $p_\perp$-spectra at various centrality parameters, assuming for the moment 
vanishing $\epsilon$ and $\delta$. The result is shown in Fig. \ref{fig:allp}.
We note that the fit works as good as for the most-central case presented in (I).
The optimum values of parameters are collected in Table \ref{tab:par}, where they are also compared
to the minimum-bias fit and the joint fit to the most central PHENIX \cite{Adcox:2001mf} 
and STAR \cite{Adler:2002qm}
data. In fact, the qualitative dependence of $\tau$ and $\rho_{\rm max}$ on $c$ is as expected: the 
larger $c$, {\em i.e.} the more peripheral collision, the smaller values of the size parameters. 
Figure \ref{fig:rt} visualizes this dependence. In Table \ref{tab:par} we 
also list the ratio of $\rho_{\rm max}/\tau$, 
and the maximum and average values of the flow parameter, $\beta$. Interestingly, 
these quantities depend very weakly on $c$. They are defined as follows: 
\begin{eqnarray}
\beta_\perp^{\rm max}&=& \frac{\rho_{\rm max}}{\sqrt{\tau^2+\rho_{\rm max}^2}}, \nonumber \\
\langle \beta_\perp \rangle &=& \int_0^{\rho_{\rm max}} r dr 
\frac{r}{\sqrt{\tau^2+r^2}}/\int_0^{\rho_{\rm max}} r dr.
\end{eqnarray}

Ideally, the dependence of the $\epsilon$  parameter on $c$ should come from the measurement
of $R_{\rm side}$ at various centralities. Then the model evaluation 
of this quantity would allow to fit independently $\epsilon(c)$ to the data. We hope to be able to proceed
in such a manner in the future. Unfortunately, no necessary experimental results are available at the moment.
In this circumstance we take a reasonable theoretical estimate for 
$\epsilon(c)$ based on Ref. \cite{Kolb:2000fh}, which leads
to $\epsilon=0.1$, $0.21$, and $0.35$ in the centrality bins $0-15\%$, $15-30\%$, and $30-60\%$,
respectively. Finally, the parameter $\delta$ is obtained by fitting the model 
predictions to the $v_2$ measurements.
Our approach includes, as described in detail in (I), the decays of resonances. The calculation
is straightforward and very similar to the one discussed in (I), although takes a much longer computer time due 
to a lower degree of symmetry. The technicalities will be presented elsewhere.
The simplified results for $v_2$ presented here include all resonances up to $m_\Delta=1.232$~GeV, and 
do not take into account three-body decays. 
 
\begin{figure}[h]
\includegraphics[width=11.5cm]{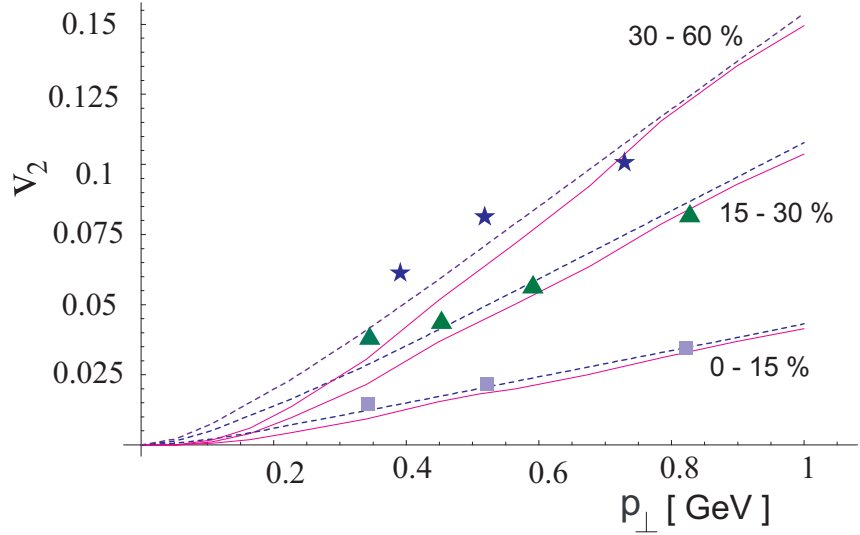} 
\caption{Dependence of $v_2$ (accumulated from all particle species) on the perpendicular momentum, $p_\perp$, for 
three centrality bins: $0-15\%$ (bottom), $15-30\%$ (middle), and $30-60\%$ (top). Thick (thin) lines 
correspond to the calculation with (without) resonance decays. Experimental points come from the PHENIX 
collaboration at $\sqrt{s_{NN}}=130~GeV$ 
\protect{\cite{Adcox:2002ms}}. The taken values for the shape-asymmetry parameter $\epsilon$ 
are, from the lowest to highest centrality bin, $0.1$, $0.21$, and $0.35$, while 
the fitted values of the flow-asymmetry parameter $\delta$ are $0.145$, $0.34$, and $0.35$, respectively. }
\label{fig:phnew}
\end{figure}

\begin{figure}[h]
\includegraphics[width=10cm]{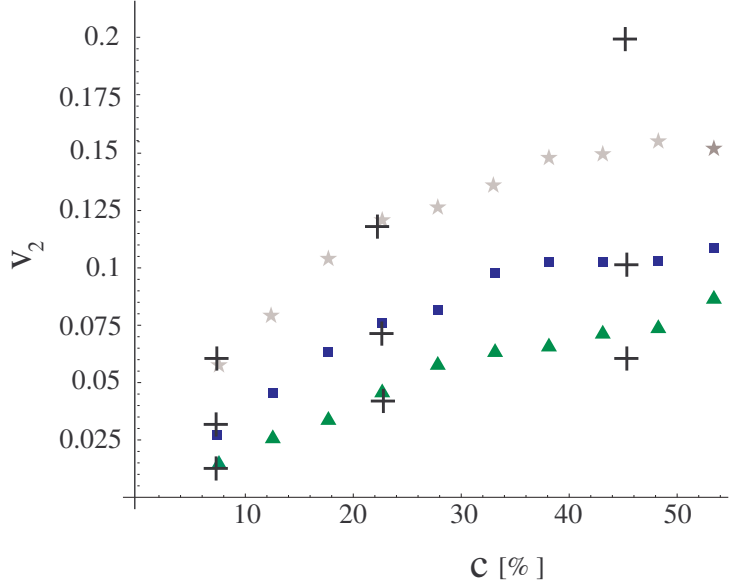} 
\caption{Dependence of $v_2$ (accumulated from all particle species) on the centrality parameter for various 
transverse-momentum bins: $0.4-0.6$~GeV (bottom), $0.6-1$~GeV (middle), and $1-2.5$~GeV (top). Crosses show 
the model calculation, while other symbols are the experimental points from the PHENIX 
collaboration at $\sqrt{s_{NN}}=130~GeV$ \cite{Adcox:2002ms}. 
The $\epsilon$ and $\delta$ parameters are the same as in Fig. \ref{fig:phnew}. }
\label{fig:cephen2}
\end{figure}

Figure \ref{fig:phnew} shows the result of our calculation for three different centrality bins. The elliptic flow 
coefficient grows with the momentum. In continues to grow for large momenta, where saturation is seen in the
experiment, however the thermal model cannot be trusted at momenta larger than about, say, 2~GeV, where 
hard dynamics is important.  
We observe that the effects of resonance decays, large in both the numerator and denominator of Eq. (\ref{v2def}),
cancel to a large degree in the ratio, and the net effect in $v_2$ is small. 
The dependence of $v_2$ on centrality for various transverse-momentum bins is show in Fig. \ref{fig:cephen2}.
We note that our calculation works well except large $c$ and $p_\perp$, {\em i.e.} except for high-momentum 
particles from most peripheral collisions.

\begin{figure}[h]
\includegraphics[width=11.5cm]{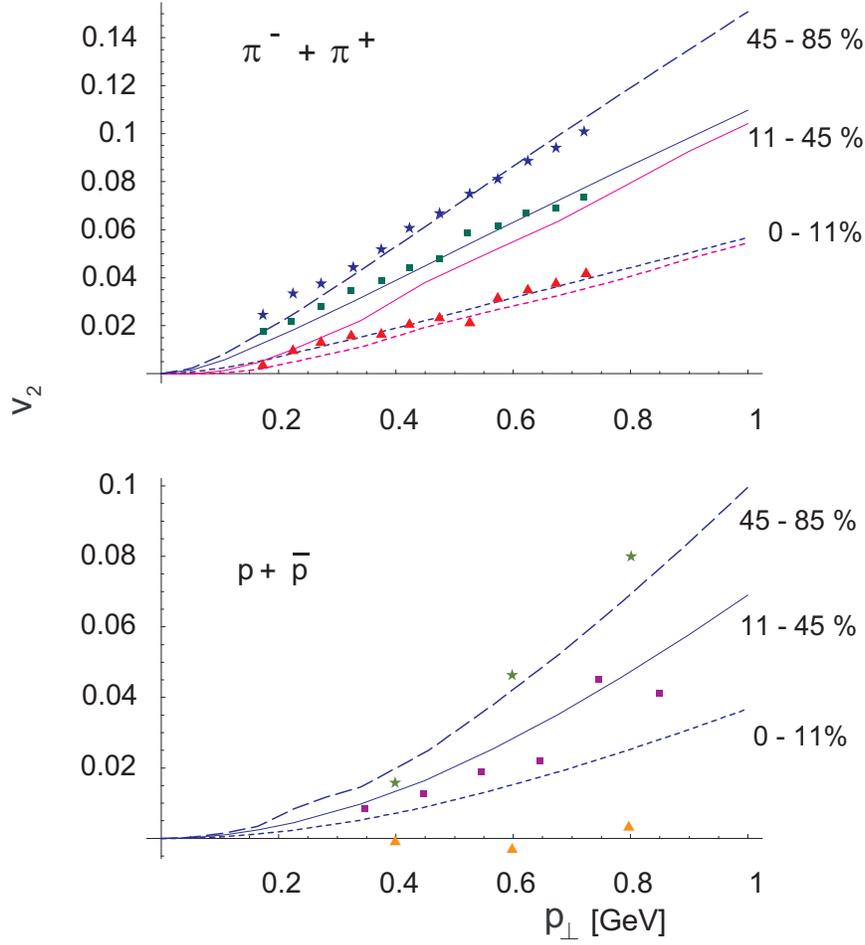} 
\caption{Dependence of $v_2$ integrated over $c$ for identified particles on the 
transverse momentum, $p_\perp$.  Thick (thin) lines 
correspond to the calculation with (without) resonance decays. Experimental points come from the 
STAR collaboration at $\sqrt{s_{NN}}=130$~GeV 
\protect{\cite{Adler:2001nb}}. The taken values for the shape-asymmetry parameter $\epsilon$ 
are, from the lowest to highest centrality bin, $0.08$, $0.25$, and $0.52$, while 
the fitted values of the flow-asymmetry parameter $\delta$ are $0.15$, $0.35$, and $0.52$, respectively.}
\label{fig:v2id}
\end{figure}

In Fig. \ref{fig:v2id} we compare the model predictions for $v_2$ integrated over $c$ for identified particles:
pions, kaons, and protons. The general characteristics, with lighter particles having larger $v_2$, are 
reproduced. The agreement is not satisfactory only for the case of protons at lowest centrality, where the data is 
compatible with zero.

To summarize this part we note that
\begin{enumerate}
\item Elliptic flow can be introduced in the thermal approach by suitably modifying the 
freeze-out hypersurface.
\item The shape-deformation parameter, $\epsilon(c)$, should and hopefully will be taken
independently from future data on the azimuthal asymmetry of the $R_{\rm side}$ HBT correlation radius
at various centralities.
\item The velocity-asymmetry parameter, $\delta(c)$, can be fitted to reproduce $v_2$. Predictions
for identified-particle $v_2$ follow and are reasonable. 
\item The $p_\perp$-dependence shows a monotonic growth, which agrees with the data 
at lower values of $p_\perp$, but certainly fails to produce saturation at large momenta, where the
thermal approach is not applicable.
\item Resonance decays do not have a very large effect on $v_2$.
\end{enumerate}

\section{HBT radii}

Now we pass to the description of the pionic Hanbury-Brown--Twiss  
correlation radii (for a review of the problem see, {\em e.g.}, 
\cite{Baym:1998ce}). The studied object is the two-particle correlation function for identical particles,
in the present case $\pi^+\pi^+$ or $\pi^-\pi^-$. It is given by    
\begin{eqnarray}
 C(\vec{q},\vec{P})=\frac{\{n_{\vec{p}_1}n_{\vec{p}_2}\}}{\{n_{\vec{p}_1}\}\{n_{\vec{p}_2}\}}, 
\end{eqnarray}
where $\{ .\}$ denotes averaging over events, $p_1$ and $p_2$ are the momenta of the pions, 
$\vec{q}=\vec{p}_2-\vec{p}_1$, and $\vec{P}=\vec{p}_1+\vec{p}_2$.
We use the Bertch-Pratt parameterization \cite{Pratt:1984su,Pratt:1986ev,Bertsch:1988db}, 
\begin{eqnarray}
 C(\vec{q},\vec{P})=1+\lambda e^{-\left ( q^2_{\rm out} R^2_{\rm out}+q^2_{\rm side} R^2_{\rm side}+ 
q^2_{\rm long} R^2_{\rm long}+ 2q_{\rm out}q_{\rm long} R^2_{\rm ol} \right ) }. 
\end{eqnarray}

First, let us briefly recall the experimental highlights. Two facts came as a great surprise with the RHIC
data. First, the $R_{\rm side}$ and $R_{\rm out}$ radii practically do not depend on the 
collision energy \cite{Adler:2001zd,Adcox:2002uc}, and acquire similar values from AGS to RHIC, despite the 
increase of the energy by almost two orders of magnitude. 
Secondly, the ratio of $R_{\rm out}$ to $R_{\rm side}$, which can be interpreted 
as a measure of the duration time of the freeze-out, is close or even less than one. This is in contradiction
to anticipations from numerous 
hydrodynamic simulations, which had predicted $R_{\rm out}/R_{\rm side}$
significantly larger than one.

Our model evaluation of the correlation function is performed 
according to the formalism of Ref. \cite{Bolz:1993hc}, with a technical approximation of 
neglecting the finite life-time of resonances. In that case
\begin{eqnarray} C(\vec{q},\vec{P}) = 1+
\frac{\left | \int d\Sigma(x) \cdot u(x) e^{i {{ q}}\cdot x} S(P \cdot u(x)) \right |^2}
{ \int d\Sigma \cdot u S({(P+{{ \frac{q}{2}}})\cdot u(x)}) 
 \int d\Sigma \cdot u S({(P-{{ \frac{q}{2}}})\cdot u(x)}) } ,
\end{eqnarray}
where the source function is 
\begin{eqnarray} 
S(p \cdot u)=\frac{1}{(2\pi)^3}e^{-(p \cdot u-\mu)/T} + {\rm 
contribution \; from \; resonances} .
\end{eqnarray}

As discussed in Ref. \cite{wbwfPRCstrange},
our model values for the geometric parameters $\tau$ and $\rho_{\rm
max}$ are low, of the order of the size of the colliding nuclei. As a result,
the values of the $R_{\rm side}$ and $R_{\rm out}$ HBT radii obtained with the procedure described above 
are about 30\% too low compared to the experiment. The problem can be
alleviated with the inclusion of the excluded-volume (Van der Waals)
corrections.  Such effects have been realized to be important since the
early thermal studies of the particle production in relativistic
heavy-ion collisions
\cite{Braun-Munzinger:1995xr,Braun-Munzinger:1996bp,Yen:1997rv,Yen:1998pa}, 
where they led to a
significant dilution of system. In the case of the Boltzmann 
statistics, which is a very good approximation in our
case \cite{mm}, the excluded volume corrections bring in a factor
\cite{Yen:1997rv}
\begin{equation}
\frac{e^{-P v_i /T}}{1+\sum_j v_j e^{-P v_j /T} n_j},
\label{vdw}
\end{equation}
into the phase-space integrals, 
where $P$ is the pressure, $v_i=4 \frac{4}{3}\pi r_i^3$ 
is the excluded volume for the 
particle of species $i$, and $n_i$ is the density of particles of species $i$.
The pressure can be calculated self-consistently from the equation 
\begin{equation}
P=\sum_i P^0_i(T, \mu_i-P v_i/T)=\sum_i P^0_i(T, \mu_i)e^{-P v_i/T},
\label{pres}
\end{equation}
where $P^0_i$ denotes the partial pressure of the ideal gas of hadrons 
of species $i$.
With the simplest assumption that the excluded volumes for all particles are equal, 
$r_i=r$, the excluded-volume correction manifests itself as a common scale factor,
which we may denote by $S^{-3}$. 
The Frye-Cooper formula can then be written in the form \cite{wbwfPRL,ZAKOPANE}
\begin{eqnarray}
\frac{dN_{i}}{d^{2}p_{\perp }dy} &=&\ \tau ^{3}\int_{-\infty }^{+\infty
}d\alpha _{\parallel }\int_{0}^{\rho _{\max }/\tau }{\rm sinh}  \alpha _{\perp
}d\left( {\rm sinh}  \alpha _{\perp }\right)  \nonumber \\
&&\times \int_{0}^{2\pi }
d\xi \, p\cdot u \, S^{-3} f_{i}\left( p\cdot u\right) ,
\label{dNi}
\end{eqnarray}
where
$p\cdot u=m_{\perp }{\rm cosh} \alpha _{\parallel } {\rm cosh}  \alpha
_{\perp }-p_{\perp }\cos \xi \, {\rm sinh}  \alpha _{\perp }$.
As can be immediately seen from this expression, the presence of the factor $S^{-3}$ in Eq. (\ref{dNi})
may be compensated by rescaling $\rho$ and $\tau$
by the factor $S$. That way the system becomes more dilute and larger in such
a way, that the {\em particle multiplicities and the spectra are left intact}.

With our values of the thermodynamic parameters $\sum_i P^0_i(T,
\mu_i)=80$~MeV/fm$^{3}$, and we find $S=1.3$ with $r=0.6$~fm.
Such a value of the excluded volume is compatible with values
typically obtained in other calculations. The increase of the
size parameters by 30\% is what we need to bring the 
values of $R_{\rm side}$ and $R_{\rm out}$ up to the experimental ball park. 

\begin{figure}[tb]
\includegraphics[width=7.5cm]{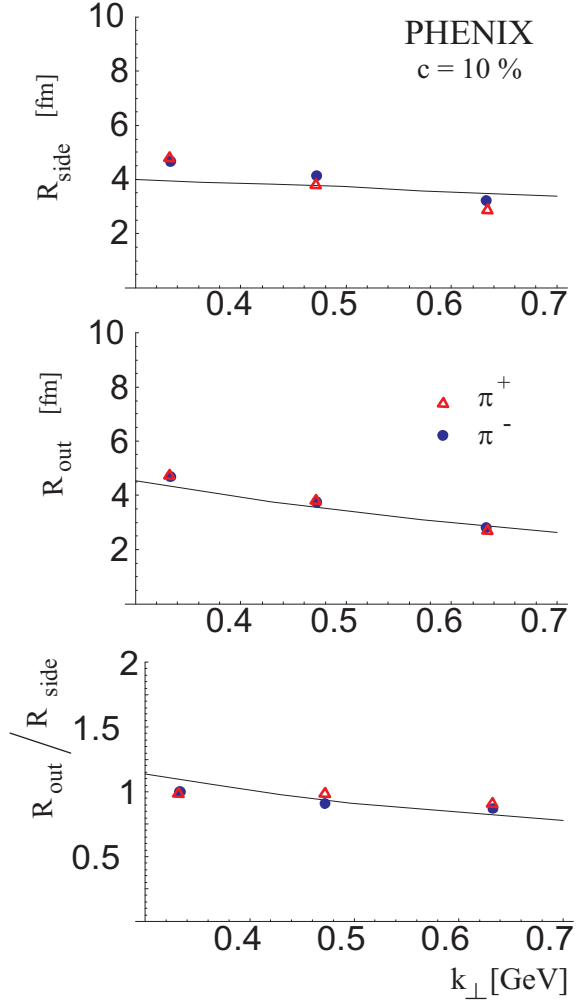} 
\caption{Model predictions for the pionic $R_{\rm side}$ and $R_{\rm out}$ HBT 
correlation radii (top two panels), and their ratio (bottom panel), 
confronted with the PHENIX data $Au+Au$ data at $\sqrt{s_{NN}}=130~{\rm GeV}$ and average 
centrality 10\%. The quantity $k_\perp$ is the total momentum of the pion pair.}
\label{fig:hbt}
\end{figure}

Our results are shown in Fig. \ref{fig:hbt}.
We note that the agreement with data is very reasonable. In particular, the 
ratio of $R_{\rm out}/R_{\rm side}$ is close to one, and drops below one
at larger values of the pair momentum $k_\perp$. The plots of $R_{\rm side}$
and $R_{\rm out}$ include the excluded-volume correction factor $S=1.3$. We observe that 
these radii decrease with the pair momentum $k_\perp$, although somewhat slower than
indicated by the data. 
We note that the radius $R_{\rm long}$ cannot be reliably evaluated in our model. 
This is due to the assumption of the 
boost invariance, {\em cf.} Eq. (I.2), which leads to too large values of $R_{\rm long}$. 

\section{Conclusion}

To conclude, we list the main results of our approach:

\begin{enumerate}
\item The thermal model works for the  particle ratios, see (I). 
\item Supplied with expansion, it works for the $p_\perp$-spectra, see (I). The 
complete treatment of resonances is essential, and the assumption of single freeze-out 
leads to good predictions. Moreover, the strange particles, including $\Omega$, are 
described properly with no need for extra parameters. 
\item Supplied with azimuthal asymmetry, the model can be used to describe 
the elliptic flow at moderate transverse momenta, up to $p_\perp \sim 2$~GeV. 
The fitted values of the velocity asymmetry parameter, $\delta$, are 
reasonable. 
\item Supplied with the excluded-volume corrections, the model works also for the HBT radii
$R_{\rm side}$ and $R_{\rm out}$. In particular, the ratio of $R_{\rm out}/R_{\rm side}$ 
is close to 1.  
\item The description is efficient, involving two thermal parameters $T$ and $\mu_B$, 
two size parameters $\tau$ and $\rho_{\rm max}$, and in the case of 
azimuthal asymmetry, two deformation parameters, $\epsilon$ and $\delta$. 
\item Finally, we note that the model also works for the case of 
RHIC at $\sqrt{s_{NN}}=200$~GeV~A as well as for SPS at $\sqrt{s_{NN}}=17$~GeV \cite{wbHIRSCHEGG,RHICvSPS}.
\end{enumerate}

\begin{theacknowledgments}
This research has been supported by PRAXIS XXI/BCC/429/94, and by the 
Polish State Committee for Scientific Research (KBN),
grants 2~P03B~11623 and 2~P03B~09419.
\end{theacknowledgments}

\bibliographystyle{aipprocl}

\bibliography{paper}

\end{document}